\documentclass[prl,twocolumn,showpacs]{revtex4}   
\usepackage{amsmath,amsfonts}
\usepackage{epsfig}
\usepackage{graphicx}
\newcommand{\la}{\lambda}
\newcommand{\om}{\omega}
\newcommand{\ep}{\epsilon}
\newcommand{\de}{\delta}
\newcommand{\bea}{\begin{eqnarray}}
\newcommand{\beq}{\begin{equation}}
\newcommand{\eea}{\end{eqnarray}}
\newcommand{\eeq}{\end{equation}}

\begin{document}
\title{Time behaviour near to spectral singularities}
\author{W.\ D.\ Heiss}
\affiliation
{National Institute for Theoretical Physics,
Stellenbosch Institute for Advanced Study, and Institute of Theoretical Physics,
University of Stellenbosch, 7602 Matieland, South Africa}
\begin{abstract} Spectral singularities such as exceptional points invoke
specific physical effects. The present paper focuses upon
the time dependent 
solutions of the Schr\"odinger equation. In a simple model
it is demonstrated that - depending on initial conditions - 
within close proximity of exceptional points
the time behaviour of the wave function displays characteristic
features such as very fast decay or the opposite, i.e.~very long
life time. At the exceptional point the wave function typically
has a linear term in time besides the usual exponential behaviour.
\end{abstract}
\pacs{
      78.47.jd,
      34.50.-s,
      02.40.Xx}

\maketitle

\section{Introduction}
Recently the effects of spectral singularities upon scattering have
been discussed in general terms \cite{hena}using a particular model
for demonstration. It is the exceptional
points (EP) \cite{kato} that generically occur in Hamiltonians of open systems
for specific (complex) values of some parameters \cite{he,gu}. The EPs
are associated with the coalescence of two (or more) eigenvalues under variation
of appropriate parameters; in contrast to a degeneracy the corresponding
eigenstates also coalesce and have zero norm. For that reason
they give rise to a double pole in the scattering function \cite{mond} and
thus to characteristic physical effects. Here we mention 
atomic and molecular physics \cite{sol,cart,lef}, optics \cite{ber1},
nuclear physics \cite{ok} and in different theoretical
context ${\cal PT}$-symmetric models \cite{zno}, to name just a few.
Depending on the particular situation EPs can signal a phase transition \cite{hege,cej}.
The topological structure being a square root branch point in the complex plane
has been shown experimentally to be a physical reality \cite{demb1,demb2,ep}.

The present paper resumes the model dealt with in \cite{hena}, however, from a
very different viewpoint. While the scattering deals with a stationary physical
situation, we here investigate the effect of EPs upon the time behaviour of solutions
of the time dependent Schr\"odinger equation. Such effects have been seen recently
in \cite{wiers} with optical micro-spirals, in \cite{darm} for Rabi oscillations and
earlier in \cite{mvb} for a singular di-electric tensor. It is expected that the rapidly
improving experimental techniques of lasers applied for instance to atomic and molecular physics
will focus upon time behaviour to an increasing extent \cite{priv}.

The following section presents the time evolution with the emphasis at
or close to an EP.
A generic physical example is given in section 3 showing the characteristic
and dramatic changes of the time dependent wave function in the proximity of EPs.
A summary concludes the paper.

\section{Time evolution}
\label{sec:1}
\subsection{The Model}
\label{sec:1a}

It is well established that, in the close vicinity of an EP, a
two-dimensional matrix model suffices to capture all essential
features associated with the singularity. We thus begin with
the model Hamiltonian
\bea
H(\la )&=& H_0+H_1(\la )=H_0+\la V \nonumber\\
&=&
\begin{pmatrix} \om_1 & 0 \\ 0 & \om_2  \end{pmatrix}
+\la \begin{pmatrix} \ep_1 & \de \\ \de & \ep_2 \end{pmatrix}
\label{ham}
\eea
where the parameters $\omega_k$ and $\epsilon_k$ determine the
non-in\-ter\-ac\-ting resonance
energies $E_k=\omega_k+\la \epsilon_k, \,k=1,2$. The time evolution of
a two dimensional state vector $|\psi (t)\rangle $ with the initial
condition  $|\psi (0)\rangle =\{C_1,C_2\}^T$ is given by
\beq
|\psi (t)\rangle = \sum _{k=1,2} \exp (-i E_k(\la )t) 
\langle \phi_k|\psi (0) \rangle |\phi _k\rangle
\label{time}
\eeq
with the eigenvalues
\bea
E_{1,2}(\la ) &=&\frac {1}{2} (\om _1+\om _2+\la (\ep _1 +\ep _1)\mp D \\
D &=& \sqrt {CC (\la -EP1)(\la -EP2)}) \\
CC &=& 4 \de ^2+(\ep _1-\ep _2)^2
\eea
expressed in terms of the EPs
\bea
EP1&=&\frac {i(\om _1-\om _2)}{-2 \de -i(\ep _1-\ep _2)}\\ 
EP2&=&\frac {i(\om _1-\om _2)}{+2 \de -i(\ep _1-\ep _2)}
\eea
and the normalised eigenvectors
\bea
|\phi _1\rangle & & = \begin{pmatrix} D+2 \la \de , \\
\om _1-\om _2+\la (\ep _1 -\ep _2)
  \end{pmatrix} \nonumber \\
& \times & \frac{1}{\sqrt{(\om _1-\om _2+\la (\ep _1 -\ep _2))^2+(2 \la \de + D)^2}}
   \eea
with $|\phi _2\rangle $ obtained from $|\phi _1\rangle $  
by the replacement $\de \to -\de $.
Note that the denominator of $|\phi _k\rangle $ vanishes when 
$\la \to EP1$ or $\la \to EP2$,
the leading order being $(\la - EP)^{1/4}$ \cite{he}.

Actually, as in \cite{hena}, we have chosen for the eigenvectors
a basis rotated by the angle
$\pi/4$ rather than the basis given by (\ref{ham}).
This specific observational basis is essential to ensure that even for the
non-interacting case ($\delta =0$) the two rotated channels feature equally and
are observed simultaneously, while - for $\delta =0$ - the states in the basis given
by (\ref{ham}) live in orthogonal spaces.  The difference between the
two rotated channels appears when $\delta $ is switched on as is discussed in
the following sections.

Inserting the explicit expressions into (\ref{time}) we obtain for
the components denoted by $z_{1,2}(t)$ of the two component solution
$$ |\psi (t)\rangle =\begin{pmatrix} z_1(t) \\ z_2(t) \end{pmatrix}$$
the result
\bea
z_2(t)=& &C_2\bigg ( \frac {\exp(-iE_1 t)+\exp(-iE_2 t)}{2} \nonumber \\
+& &\de \frac{\exp(-iE_1 t)-\exp(-iE_2 t)}{ D} \bigg )\nonumber  \\
+\; C_1& & \frac {\exp(-iE_1 t)-\exp(-iE_2 t)}{2 D}\nonumber  \\
\times & &  (\om _1-\om _2+\la (\ep _1 -\ep _2)). \label{slt}
\eea
The first component $z_1(t)$ is obtained from $z_2(t)$ by the
replacements $C_1\leftrightarrow C_2 $ and $\de \to -\de $.

Before discussing the detailed behaviour in an actual physical
situation we first turn to the analytic solution at an EP.

\subsection{Time behaviour at an exceptional point}
\label{sec:2}

When $\la $ approaches an EP the denominator $D$ in (\ref{slt}) vanishes.
So do the corresponding numerators since $E_1 \to E_2$ when $\la \to EP$.
A finite limit is obtained and reads
\bea
& &z_2^{EP1}(t)=       \nonumber        \\
& & \frac {(\ep_1-\ep_2-2i\de +i\de (\om_1-\om_2)t)C_2-
 \de (\om_1-\om_2)t C_1}{\ep_1-\ep_2-2 i\de} \nonumber \\
& & \times  \exp(-it\frac{i(\ep_1 \om_2-\ep_2
  \om_1)+\de (\om_1+\om_2)}
{i(\ep_1-\ep_2)+2\de } )   \label{z2EP} 
\eea
and 
\bea
& &z_1^{EP1}(t)=        \nonumber       \\
& & \frac {(\ep_1-\ep_2-2i\de -i\de (\om_1-\om_2)t)C_1-
 \de (\om_1-\om_2)t C_2}{\ep_1-\ep_2-2 i\de} \nonumber \\
& & \times  \exp(-it\frac{i(\ep_1 \om_2-\ep_2
  \om_1)+\de (\om_1+\om_2)}
{i(\ep_1-\ep_2)+2\de } ). \label{z1EP} 
\eea

The essential point is the linear time dependence occurring in the
coefficient of the exponentials. The 'singularity', usually associated
with an EP, invokes in the time frame the additional linear time
dependence. This is of course no surprise; it is a well known result from
the theory of ordinary differential equations. The evolution equation
$i\partial _t|\psi (t)\rangle = H|\psi(t)$ gives rise to linear terms
in time if $H$ cannot be diagonalised but has a non-diagonal
Jordan normal form; this is precisely the case at an EP. It is here
where the associate vector defined by $$(H(\la =EP)-E_{EP})|\phi_{\rm assoc}\rangle
=|\phi_{EP}\rangle $$ plays its role (see for instance \cite{sey}). For more details and
generalisations (higher order terms in $t$) we refer to the Appendix.

We note that the linear time dependence at an EP has been noticed in \cite{mvb}
for a non-diagonalisable di-electric tensor, and recently in
\cite{wiers} for asymmetric scattering in optical micro-spirals,
and also indirectly in \cite{darm} for Rabi
oscillations in a microwave cavity.

We mention that these results cannot be obtained easily
from a simple Fourier transform of the corresponding Green's function
$G(E)=(E-H)^{-1}$ as given in \cite{hena}. The reason is due to the fact
that (i) the residues blow up owing to the vanishing norm of the
eigenfunctions at the EP (actually $G(E)$ has a pole
of second order \cite{mond}), and (ii) the associate vector does not naturally occur.
Of course, the mathematical connection does exist, yet the results in
the time domain are much easier obtained by the approach used in the
present paper.

\section{A particular physical case}
For illustration we use the same parameters as in \cite{hena},
that is
$\omega_1=1.55-0.007i,\,\omega_2=1.1-0.007i,\,
\epsilon_1=-0.4-0.0006i,\,\epsilon_2=0.4+0.0005i$
and $\de =0.0115 i$.
For $\la $ we restrict ourselves to the interval [0.53,0.59]; the
critical value is $\la _c=0.563$, it is there where a resonance pole
comes closest to the real axis. The close proximity of two exceptional
points makes the movement of the resonance pole under variation of
$\la $ rather swift and dramatic. This is illustrated in Figure 1 where
trajectories are drawn in the complex energy plane.

\begin{figure}
\resizebox{0.40\textwidth}{!}{%
\includegraphics{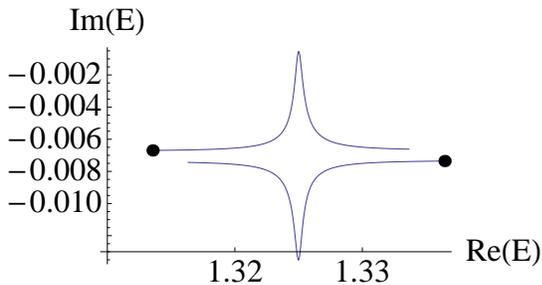}
}
\caption{Energy eigenvalue trajectories when $\la $ sweeps from 0.53
  to 0.59.  Note that for our choice of parameters the
trajectories run in opposite directions, the respective starting
points are indicated by a dot.}
\label{traj}
\end{figure}

The specific values chosen have no particular significance, they serve to illustrate the
principle, that is the effect of a near EP upon the behaviour of
resonance poles and thus the time behaviour of
the state vectors; any other
set that invokes repulsion and coalescence of resonances would serve
the same purpose as long as the imaginary parts of the poles are near to
the real axis. While we did not specify units for the present
illustration, choosing for instance meV for the energies would imply
picoseconds as time units in Figures 2 and 3.

\begin{figure}[b]
\resizebox{0.47\textwidth}{!}{%
\includegraphics{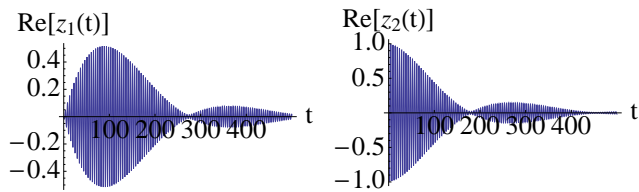}
}
\caption{Time dependence of the two components of the wave function at
$\la =0.53$. The corresponding imaginary parts look the same.}
\label{bef}
\end{figure}

Similar in spirit to the procedure in \cite{wiers} we choose for the
initial condition of $\psi (t)$ a specific polarisation and consider
the two cases $\psi (0)=\{1,0\}^T$ and $\psi (0)=\{0,1\}^T$. In 
Figure 2 the two components are plotted for the latter case at a value of $\la $
where the two resonances are well separated (the dots in Figure 1). Three
main features become obvious: (i) the initial conditions require that
the first and second component begin at zero and unity, respectively;
(ii) the beat is determined by the frequency difference of the two
near resonances while the coupling invokes an amplitude transfer
between the two components; (iii) the damping -- not to be confused
with the beat -- is determined by the width (the imaginary part) of the two
resonances, $\exp (-t\Gamma)$  gives the envelope of the damped
oscillation. Note that for our choice of parameters we find $\Gamma=0.007$ for the width
while the beat frequency $\Delta E=\Re [E_1]-\Re [E_2]=0.025$ is 
larger than the width (a few beats can be accommodated under the
envelope). Swapping the initial conditions would essentially swap the
two components.

A dramatic change occurs when $\psi (t)$ is plotted at the critical
point $\la _c=0.563$ being the point where the trajectories peak in
Figure 1. Moreover, the change affects the two different initial
conditions in a very different and characteristic way.
Three important changes are noticed in Figure 3: (i) the beat has disappeared;
(ii) for the initial condition $\psi (0)=\{0,1\}^T$ (top row) the
leading component is very weakly damped while the other component
remains rather small, yet it is also weakly damped; (iii) for the initial
condition $\psi (0)=\{1,0\}^T$ (bottom row) the leading component
$z_1(t)$ is strongly damped while the other component $z_2(t)$ is
essentially like that for the
top row (here $z_1(t)$). Again we emphasise that the respective
envelopes of $z_2(t)$ (top) and $z_1(t)$ (bottom) are given by the
damping related to the widths of the top and bottom peak in Figure 1,
the values are $\Gamma _{\rm top}=0.0005$ and $\Gamma _{\rm bot}=0.013$.

\begin{figure}[b]
\resizebox{0.47\textwidth}{!}{%
\includegraphics{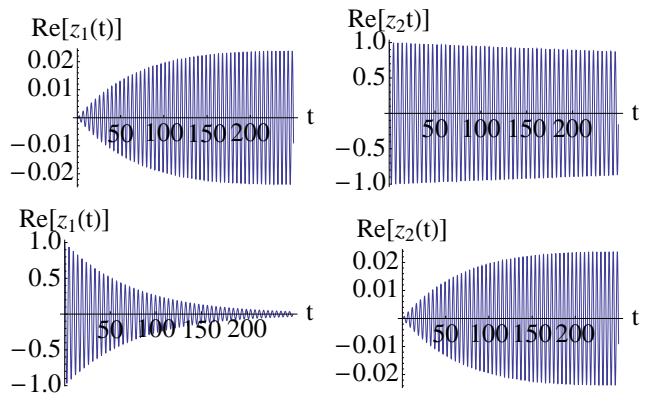}
}
\caption{Time dependence of the two components of the wave function at
the critical points $\la =0.563$. The top row is with initial
conditions as in Figure 2 while in the second row the initial components
are interchanged. Note that the time axis extends only half as far as
the one in Figure 2.}
\label{at}
\end{figure}

These findings are explained using Figure 1 and
(\ref{slt},\ref{z2EP}) and (\ref{z1EP}).
At the critical point there is one narrow resonance (top peak in
Figure 1)
and one broader resonance (bottom peak in Figure 1). As the frequencies
(the real parts of the energy) are the same, there is no beat. Now we
first turn to $z_2(t)$ of the top row and $z_1(t)$ of the bottom row
in Figure 3. 
As seen from (\ref{slt}) the expression associated with $C_2=1$ has
two terms: the sum of the exponentials with the two eigenenergies (and the
factor 1/2) and their difference (with the factor $\de /D$); the two
terms are added. Similarly, the corresponding two terms for $z_1(t)$
being associated with
$C_1=1,\,C_2=0$ are subtracted. The conspiring behaviour due to the
sum in the one case and the difference in the other
becomes evident in
(\ref{z2EP}) (with $C_2=1,\,C_1=0$) and (\ref{z1EP}) (with
$C_1=1,\,C_2=0$), respectively. The energies in the exponential are of
course different in (\ref{slt}), there is the small and large damping
(width). When adding the two terms as for $z_2(t)$, top row of Figure 3, the small
damping prevails while subtracting as for $z_1(t)$, bottom row, the
large damping is dominant. 
At the top row $z_1(t)$ is immaterial in comparison
with $z_2(t)$, while at the bottom row $z_2(t)$ becomes stronger than
$z_1(t)$. We return to this interesting aspect in the last section.

These rather dramatic differences for the time
beha\-viour depending on the initial conditions are expected to be observable
as a physical effect; in fact, the polarisation in \cite{wiers} is
just one case in point. The findings discussed in the present paper
complement beautifully the results presented in \cite{hena}. This is
discussed in greater detail in the following section.

When going beyond the critical point with $\la $, say at $\la =0.59$, Figure 2 is
essentially repeated. The graphs for the imaginary parts of the complex components
$z_{1,2}(t)$ look essentially as the ones given.

\section{Summary and Conclusion}
In \cite{hena} the effect of EPs has been discussed within the same
model and parameters as in the present paper,
but the emphasis was on resonance scattering,
that is a stationary physical situation. Now we focus
upon the related time dependent problem, that is the solution of the
time dependent Schr\"odinger equation, which is formally given by
$$|\psi (t) \rangle=\exp (-iHt)|\psi (0)\rangle .$$
For the purpose of the present paper the Hamiltonian is modelled by a
$2\times 2$-matrix describing an open system. It depends on a
parameter $\la $ and of interest is the behaviour of
$|\psi (t)\rangle $ when $\la $ is at an EP or sweeping over a range nearby an
EP. Similar to the findings in \cite{hena} the variation of $\la $
causes dramatic changes due to the proximity of EPs.

When the resonances are well separated, the time behaviour is damped
owing to the widths of the eigenstates, but there is also a beat owing
to the difference of the two frequencies. This changes dramatically when a
pair of EPs is approached: the beat disappears, and depending on the
initial condition the leading component is very weakly or very
strongly damped. This is reminiscent of the very sharp resonance peak
in the one channel and a broad peak in the other channel in two
channel scattering \cite{hena}. In addition, while the smaller component
remains insignificant for the weakly damped case, it becomes the
stronger component for the strongly damped case. In fact, while there the
leading component tends quickly to zero due to the larger width, the associated
smaller component becomes dominant and is actually weakly damped. This
corresponds exactly to the {\it peak on the broad peak} as detected in \cite{hena}.

As a general result we noticed that the singular be\-ha\-vi\-our usually associated
with EPs leads to a much milder pattern in the time frame. The
'self-orthogonality' of the eigenstates appears in a less dramatic
way: depending on the order of the EP a polynomial in the time
variable occurs besides the exponentials; in the simplest case it is just a
linear term \cite{wiers,darm,mvb}. 

As an increased activity of experiments using time behaviour in atomic and molecular
physics can be expected, the findings of the present paper may be of special relevance.

\section*{Appendix}
This is a rehash of known facts from linear ordinary differential equations.
If the time independent square matrix operator $\cal O$ is diagonalisable, 
the evolution equation
$$\frac {\rm d}{\rm d t}|\chi(t) \rangle ={\cal O}|\chi (t) \rangle $$
with the initial condition $|\chi (0)\rangle $ is solved by 
$$|\chi (t)\rangle =  \sum _k \exp (E_kt) 
\frac{\langle \phi_k^l|\chi (0) \rangle |\phi _k^r\rangle }
{\langle \phi _k^l|\phi _k^r\rangle }$$
with $E_k$ and $|\phi _k^r\rangle $  and $\langle \phi _k^l| $
forming the right and left hand eigensystem
of the operator $\cal O$. At an EP, when two eigenvalues coalesce,
the matrix cannot be diagonalised but has the Jordan decomposition
$${\cal O}=S\,J\,S^{-1}$$
with
$$J=\begin{pmatrix} E_{EP} & 1 & & & \\
0 & E_{EP} & & 0 & \\
& & E_3 & & \\
 & 0 & & \ddots   & \\
& & & & E_N  \end{pmatrix}$$
where we have listed the two EPs first. Now the solution of
$$\frac {\rm d}{\rm d t}|\xi(t) \rangle =J\,|\xi (t) \rangle $$
reads for the initial vector $\{C_1,\ldots,C_N\}^T$
\bea
|\xi (t)\rangle =&\{& (C_1+tC_2)\exp (E_{EP}t),C_2\exp (E_{EP}t),\nonumber \\
& &C_3\exp (E_3t),\ldots,C_N\exp (E_Nt)\}^T  \nonumber
\eea
which can be transformed back into the basis of the original $\cal O$
using the similarity transformation $S$. A so-called associate vector
related to the eigenvector $|\phi _{EP}\rangle $ 
by   $$({\cal O}-E_{EP})|\phi_{\rm assoc}\rangle=|\phi_{EP}\rangle $$
features here in the second column of $S$. For the special case of Section 2
it can be chosen as in the second column of $S$ which reads
$$
S=\begin{pmatrix} i & \frac{2 \delta +i(\ep_1-\ep_2)}{\delta (\om_1-\om_2)} \\
1 & 0  \end{pmatrix}.
$$
This explicit form explains the terms linear in $t$ occurring in (\ref{z2EP}) and (\ref{z1EP}).

It is clear how to generalise this when more than two levels coalesce.
For $k$ coalescing levels the first component of $|\xi (t)\rangle $ is
given by $$(C_1+t\,C_2+t^2/2!\,C_3+\ldots +t^{k-1}/(k-1)!\,C_k)\exp (E_{EP}t)$$
and accordingly for the following components.

%

\begin{thebibliography}{}

\bibitem{hena}
                     W.D. Heiss and R.G. Nazmitdinov
                     Eur.Phys.J. D\textbf{58}, 53 (2010).

\bibitem{kato}       T.~Kato, \emph{Perturbation theory of linear operators}
                     (Springer, Berlin, 1966).


\bibitem{he}
                     W.D. Heiss,
                     Eur.Phys.J. D\textbf{7}, 1 (1999);
                     Phys.Rev. E\textbf{61}, 929 (2000);
                     Czech.J.Phys. \textbf{54}, 1091 (2004).

\bibitem{gu}         U. G\"unther, I. Rotter, B.F. Samsonov,
                     J.Phys.A: Math.Theor. \textbf{40}, 8815 (2007).

\bibitem{mond}
                    E.~Hernandez, A.~Jaureguit and A.~Mondragon,  
                    J. Phys.~A {\bf 33}, 4507 (2000) 

\bibitem{sol}
                     E.A. Solov'ev,
                     Sov.Phys.Usp. \textbf{32}, 228 (1989).

\bibitem{cart}

                     H. Cartarius, J. Main, G. Wunner,
                     Phys. Rev. A\textbf{79}, 053408 (2009).

\bibitem{lef}        R. Lefebvre {\it etal.}, 
                     Phys.Rev.Lett.  \textbf{103}, 123003 (2009).


\bibitem{ber1}
                     M.V. Berry, D.H.J. O'Dell,
                     J.Phys.A\textbf{31}, 2093 (1998).

\bibitem{ok}
                     J. Okolowicz, M. Ploczajczak,
                     Phys.Rev. C\textbf{80}, 034619 (2009).

\bibitem{zno}

                     M. Znojil,
                     Phys.Lett. B\textbf{647}, 225 (2007).

\bibitem{hege}

                      W.D. Heiss, F.G. Scholtz, H.B. Geyer,
                     J.Phys.A: Math.Gen. \textbf{38}, 1843 (2005).

\bibitem{cej}

                     P. Cejnar, S. Heinze, M. Macek,
                     Phys.Rev.Lett. \textbf{99}, 100601 (2007).

\bibitem{demb1}

                     C. Dembowski, H.-D. Gr\"{a}f, H.L. Harney, A. Heine,
                     W.D. Heiss, H. Rehfeld, A. Richter,
                     Phys.Rev.Lett. \textbf{86}, 787 (2001).
\bibitem{demb2}

                     C. Dembowski, B. Dietz, H.-D. Gr\"{a}f,
                     H.L. Harney, A. Heine, W.D. Heiss, A. Richter,
                     Phys.Rev. E\textbf{69}, 056216 (2004).

\bibitem{ep}

                     S.-B. Lee,  J. Yang, S. Moon, S.-Y. Lee, J.-B. Shim,
                     S.W. Kim, J.-H. Lee, K. An,
                     Phys.Rev.Lett. \textbf{103}, 134101 (2009).

                    

\bibitem{wiers}     J. Wiersig, S.W. Kim and M. Hentschel,
                    Phys.Rev. A \textbf{78}, 53809 (2008).

\bibitem{darm}       B. Dietz {\it etal.}
                     Phys. Rev. E\textbf{75}, 027201 (2007).

\bibitem{mvb}        M.V.~Berry, Current Science {\bf 67}, 220 (1994).

\bibitem{priv}       Heinrich Schwoerer (Laser Centre, Stellenbosch), private communication


\bibitem{sey}
                     A.A. Mailybaev, O.N. Kirillov, A.P. Seyranian,
                     Phys.Rev. A\textbf{72}, 014104 (2005).



\end{thebibliography}
%

\end{document}